\documentclass{sf2a-conf2015}
\usepackage{graphicx}
\usepackage{hyperref}
\usepackage[]{natbib}  
\usepackage{epstopdf}

\def\BibTeX{{\rm B\kern-.05em{\sc i\kern-.025em b}\kern-.08em
    T\kern-.1667em\lower.7ex\hbox{E}\kern-.125emX}}
\def\teff{$T_{\rm eff}$}
\bibpunct{(}{)}{;}{a}{}{,}  %%%%%%%%%%%%%  A&A bibliography style
\begin{document}

\TitreGlobal{SF2A 2015}

%%-----------------------------------------------------------------
%%      the top matter
%%

\title{The peculiar abundance pattern of the new Hg-Mn star HD 30085}

\runningtitle{HD 30085}

\author{R.Monier$^{1,}$}\address{LESIA, UMR 8109, Observatoire de Paris Meudon, Place J.Janssen, Meudon, France}\address{Lagrange, UMR 7293, Universite de Nice Sophia, Nice, France}
\author{M.Gebran}\address{Department of Physics and Astronomy, Notre Dame University - Louaize, PO Box 72, Zouk Mikael, Lebanon}
\author{F.Royer}\address{GEPI, UMR 8111, Observatoire de Paris Meudon, Place J.Janssen, Meudon, France}
\author{R.E.M.Griffin}\address{Dominion Astrophysical Observatory, 5071 West Saanich Road, Victoria, BC, V9E 2E7, Canada}

%% IF Author3 has two affiliations, the one of Author1 and a second one:
%\author{C.\,E. Author3$^{1,}$}\address{Dept. of Chess, University of Games, 35101 Las Vegas, Monaco} 

%% Keep this line, even if the page will be settled afterwards.
\setcounter{page}{237}

%%-----------------------------------------------------------------

\maketitle

%%-----------------------------------------------------------------
%%        The abstract
%% 
%%  Warning!  within the abstract:
%%  - do not use macros. 
%%  - do not use commands like: \cite, \citet, \citep ... etc.

\begin{abstract} Using high-dispersion, high-quality spectra of HD 30085
obtained with the echelle spectrograph SOPHIE at l'Observatoire de Haute
Provence, we show that this star contains strong lines of the $s$-process
elements Sr~{\sc ii}, Y~{\sc ii} and Zr~{\sc ii}. Line syntheses of the lines
yield large overabundances of Sr, Y, Zr which are characteristic of HgMn
stars. The Sr-Y-Zr triad of abundances is inverted in HD 30085 compared to that
in our solar system. The violation of the odd-even rule suggests that physical
processes such as radiative diffusion, chemical fractionation and others must
be at work in the atmosphere of HD 30085, and that the atmosphere is stable
enough to sustain them.  \end{abstract}

%% Insert the keywords (to appear in the ADS indexing)
%% Keywords must be separated by a comma

%%-----------------------------------------------------------------

\section{Introduction}
%%---------------------

HD~30085, currently assigned a spectral type of A0\,IV, is one of the 47
northern slowly-rotating early-A stars stars studied by \cite{Royer14}.  It
shows strong lines of Mn~{\sc ii} and Hg~{\sc ii}, and recently
\citet{Monier15} synthesized several lines of Mn~{\sc ii}, Fe~{\sc ii} and
Hg~{\sc ii} which are present in spectra observed with SOPHIE, using model
atmospheres and spectrum synthesis that include hyperfine structure of various
isotopes where relevant.  The synthetic spectra were adjusted iteratively to
the observed high-resolution, high signal-to-noise spectra in order to derive
the abundances of those elements.  The analysis yielded over-abundances of 40
times solar for Mn and 32000 times solar for Hg, thus demonstrating
unquestionably that the star needs to be re-classified as an HgMn star. In this
paper we focus on lines of Sr, Y, Zr which are also strong in the spectrum of
HD~30085, and derive the element abundances.
  
\section{Observations and reduction}

HD~30085 was observed twice at l'Observatoire de Haute Provence in February
2012 and December 2013, using the high-resolution mode ($R$ =75000) of SOPHIE.
Three 15-minute exposures were obtained in February 2012 and coadded to create
a mean spectrum with a $\frac{S}{N}$ ratio of about 316. A single 20-minute
exposure was acquired in December 2013, with a $\frac{S}{N}$ of $\sim$300.

\section{Lines of Sr~{\sc ii}, Zr~{\sc ii} and Y~{\sc ii} in HD 30085}

\begin{figure}[ht!]
 \centering
\includegraphics[width=0.48\textwidth,clip]{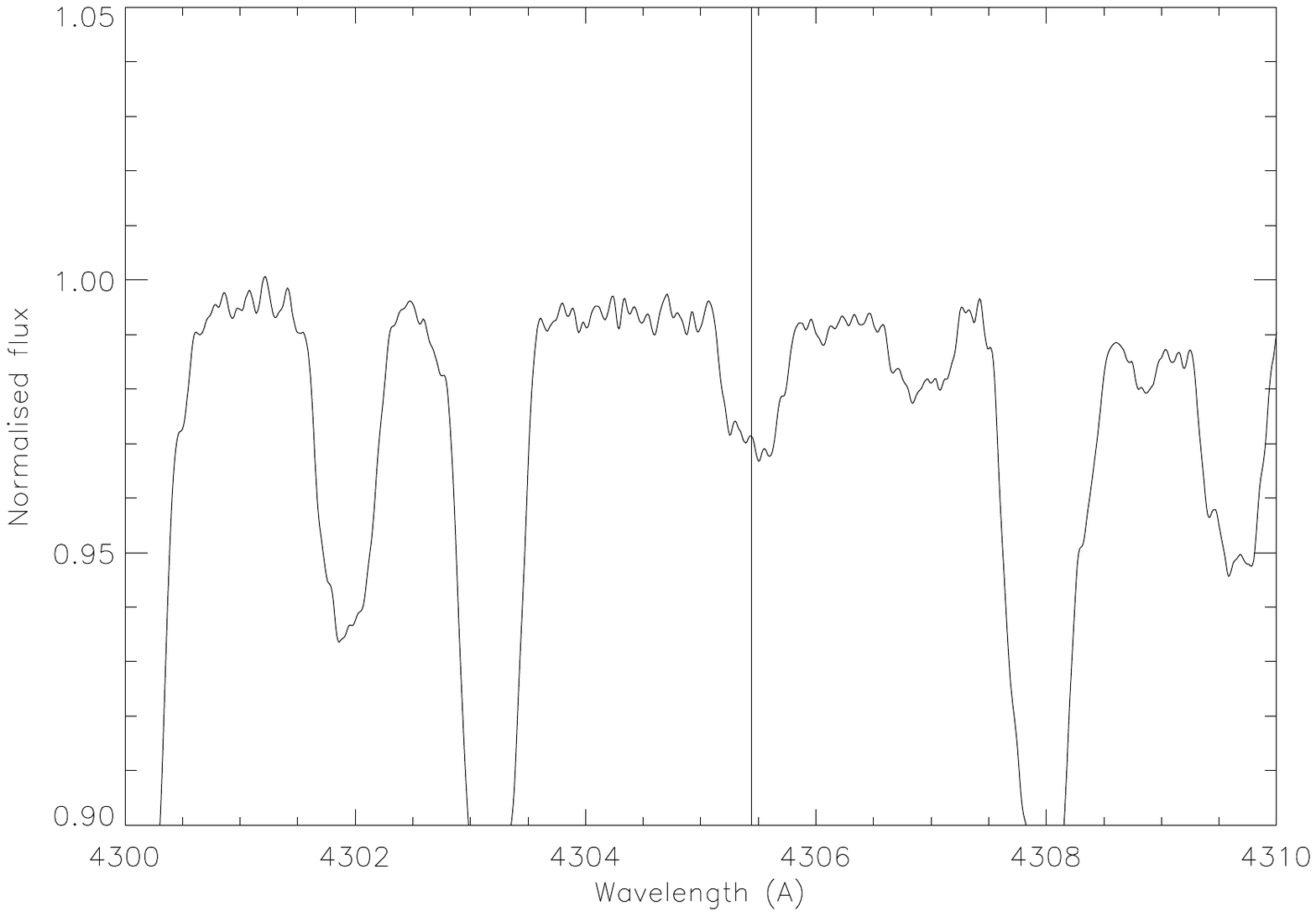}     
\includegraphics[width=0.48\textwidth,clip]{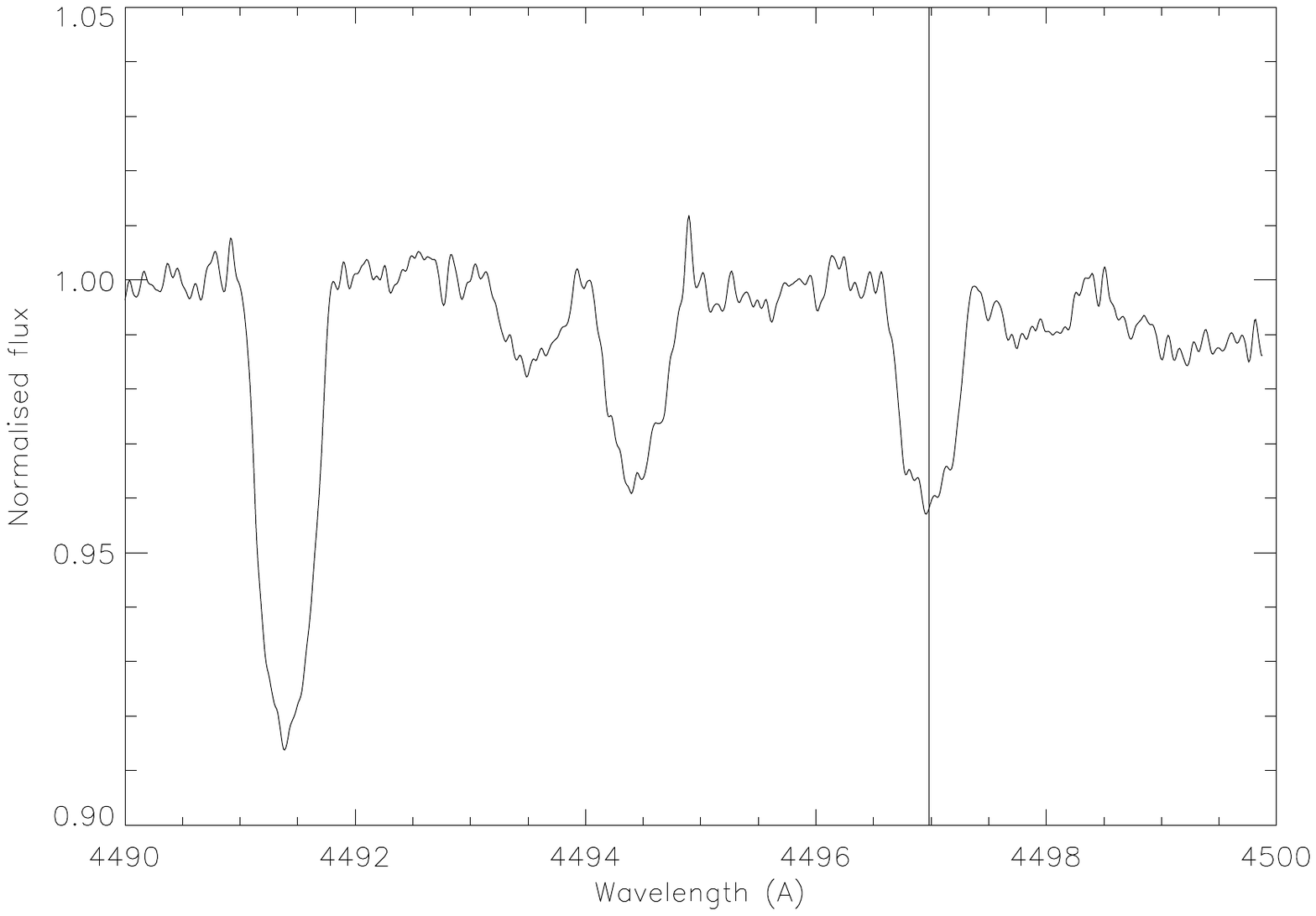} \vskip -5cm     
 \caption{{\textbf Left:} Sr {\sc ii} line at 4305 \AA\ (left). {\textbf Right:} Zr {\sc ii} line  at 4496 \AA\ (right). }
 \label{fig1}
\end{figure}

The strongest lines of Sr~{\sc ii}, Y~{\sc ii} and Zr~{\sc ii} in our line
catalogue are conspicuous in the SOPHIE spectra of HD~30085. They are listed in
Table~1 along with the measured equivalent width and derived abundance for each
transition. Only a few of these lines are unblended; most of the blends are
with lines of Cr~{\sc ii}, Mn~{\sc ii} and Fe~{\sc ii}, whose abundances were
derived in \cite{Monier15}. Fig.~\ref{fig1} displays the resonance-line profile
of Sr~{\sc ii} at 4305 \AA\ and that of Zr~{\sc ii} at 4496 \AA\ to illustrate
their strengths and the overabundances of those species.

%------------------------------------------------ Old table ------------------------------------------------
\begin{table}
\centering
\begin{tabular}{|l|l|l|l|l|p{3cm}|}
 \hline
  \multicolumn{6}{|c|}{\textbf{Lines used for abundance analysis}}   \\ \hline
  \multicolumn{1}{|c|}{Wavelengths (\AA)} &
  \multicolumn{1}{|c|}{Identification} &
  \multicolumn{1}{|c|}{Multiplet} &
  \multicolumn{1}{|c|}{EW} &
  \multicolumn{1}{|c|}{Abundance}  &
   \multicolumn{1}{|c|}{Comment}        \\ \hline
4077.71  & Sr {\sc ii} & M 2 & 74.0 & 40 $\odot$ & blend \\
4161.80 &  Sr {\sc ii} & M 3 & 11.0 & 40 $\odot$ & \\
4215.52  & Sr {\sc ii} & M 2 & 64.7 & 40 $\odot$ & blend \\
4305.45  & Sr {\sc ii} & M 3 & 14.6 & 40 $\odot$ & \\
\hline
4177.53  & Y {\sc ii}  &    & 46.0  & 300 $\odot$ & blend \\
4235.73  & Y {\sc ii}  &    & 18.3  & 300 $\odot$ &  \\
4309.63  & Y {\sc ii}  &    & 42.1  & 250 $\odot$ & blend \\ 
4358.73  & Y {\sc ii}  &    & 31.8  & 300 $\odot$ & blend \\
4398.01  & Y {\sc ii}  &    & 49.6  & 250 $\odot$ & blend \\
4422.59  & Y {\sc ii}  &    & 32.9  & 500 $\odot$ & blend \\
4682.32  & Y {\sc ii}  &    & 15.1  & 300 $\odot$ & blend \\
4823.30  & Y {\sc ii}  &    & 16.1  & 275 $\odot$ & blend \\
4883.68  & Y {\sc ii}  &    & 50.9  & 500 $\odot$ & \\
4900.12  & Y {\sc ii}  &    & 49.4  & 500 $\odot$ & blend \\
5205.72  & Y {\sc ii}  &    & 48.2  & 500 $\odot$ &  \\
5497.41  & Y {\sc ii}  &    & 31.4  & 500 $\odot$ & blend \\
5662.93  & Y {\sc ii}  &    & 51.2  & 500 $\odot$ &  \\
\hline
4443.00 & Zr {\sc ii} &     & 27.4  & 200 $\odot$ &     \\
4457.43 & Zr {\sc ii} &     & 10.0  & 100 $\odot$ &     \\
4496.98 & Zr {\sc ii} &     & 22.2  & 200 $\odot$ &     \\
5112.30 & Zr {\sc ii} &     & 12.6  & 150 $\odot$ &     \\
\hline
\end{tabular}
\caption{The strongest lines of Sr {\sc ii}, Y {\sc ii} and Zr {\sc ii} in HD 30085}  
\end{table}  
  %-----------------------------------------------

%------------------------------------------------

% classification figures here
%%
%% Example of two figures side by side
%%

\section{Model atmospheres and spectrum synthesis }

The effective temperature \teff~and surface gravity log\,$g$ of HD~30085 were
first evaluated using Napiwotzky et al's (1993) {\it uvbybeta} calibration of
Stromgren's photometry.  The derived values were \teff~= 11300 $\pm$ 200 K,
log\,$g$ = 3.95 $\pm$ 0.25.

\begin{figure}[ht!]
 \centering
\includegraphics[width=0.8\textwidth,clip]{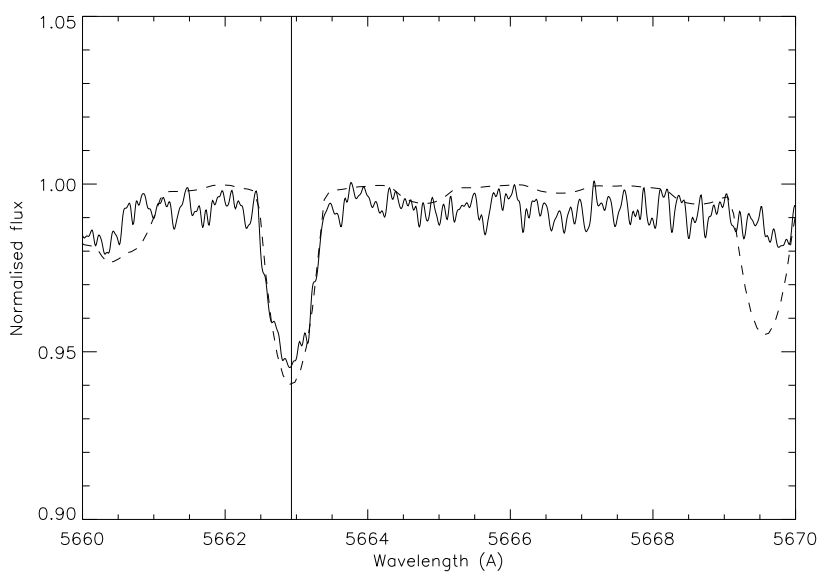}  \vskip -8.5cm    %% Note the ABSENCE of the extension .pdf , .eps or .ps  !
  \caption{Synthesis of the Y {\sc ii} line at 5662 \AA\  (observed: thick line, synthetic: dashed lines for a 500 $\odot$ overabundance)}
  \label{fig2}
\end{figure}

First a plane-parallel model atmosphere assuming radiative equilibrium and
hydrostatic equilibrium was computed using the ATLAS9 code \citep{Kurucz92},
but with the linux version that uses the new ODFs maintained by F. Castelli on
her website\footnote{http://www.oact.inaf.it/castelli/}. A line-list was built
by starting from Kurucz's (1992) ``gfhyperall.dat" file
\footnote{http://kurucz.harvard.edu/linelists/}, which includes hyperfine
splitting levels, and was upgraded by appealing to the NIST Atomic Spectra
Database \footnote{http://physics.nist.gov/cgi-bin/AtData/qlinesform} and the
VALD database operated at Uppsala University
\citep{kupka2000}\footnote{http://vald.astro.uu.se/~vald/php/vald.php}.  A grid
of synthetic spectra was then computed with SYNSPEC48 \citep{Hubeny92},
specifically to model the Sr~{\sc ii}, Y~{\sc ii} and Zr~{\sc ii} lines.  We
adopted a projected rotational velocity $v_{e} \sin i = 26$ km\,s$^{-1}$ and a
radial velocity $v_{rad} = 8.20$ km.s$^{-1}$ from \cite{Royer14}.  In
Fig.~\ref{fig2}, the observed line-profile of Y {\sc ii} at 5662.93 \AA\ is
compared with the synthetic one computed for an overabundance of Yttrium of 500
$\odot$; that overabundance provided the best fit to the observed profile.

%------------------------------------------------

%------------------------------------------------

\section{Evidence for Sr-Y-Zr excesses}

The abundances of Strontium, Yttrium and Zirconium that were derived from each analysed transition
are listed in Table~1.  The four lines of Sr~{\sc ii} yielded a consistent
overabundance of 40 $\odot$. In contrast, the Y~{\sc ii} lines yielded
overabundances ranging from 250 $\odot$ to 500 $\odot$, the scatter in the
values probably reflecting inacuracies in the atomic data of those
elements. Similarly, the Zr {\sc ii} lines yielded overabundances ranging from
100 to 200 $\odot$. We thus find that Y is more abundant than Sr and Zr in
HD~30085, which is the opposite of what is observed in the solar system.

% figure Y {\sc ii}

\section{Conclusions}

% move to conclusion following text

Lines of Sr~{\sc ii}, Y~{\sc ii} and Zr~{\sc ii} are enhanced in HD~30085. Line
synthesis reveals large overabundances in these $s$-process elements, Y being
more abundant than Sr and Zr. This violation of the odd-even rule shows that
the Sr-Y-Zr triad of abundances is inverted in HD~30085 compared to the solar
system pattern.  It strongly suggests that physical processes such as radiative
diffusion and chemical fractionation are at work in the atmosphere of HD~30085,
and that its atmosphere is stable enough for long enough to sustain such
processes.  Sr, Y and Zr are of interest for nucleosynthetic studies because
they comprise the first blocking place in the neutron absorption cross-section
for $s$-process syntheses of heavy elements in red giants.  We conclude that
HD~30085 has overabundances of Sr, Y, Zr which are characteristic of an Hg-Mn
star.  A detailed abundance analysis of other elements in this star is
currently in progress in order to complement the abundances reported here.

% Optional acknowledgements
% -------------------------
\begin{acknowledgements}
The authors acknowledge very efficient support from the night assistants at Observatoire de Haute Provence. They have used the NIST Atomic Spectra Database and the VALD database operated at Uppsala University (Kupka et al., 2000) to upgrade atomic data.
\end{acknowledgements}

%%-----------------------------
%%   Bibliography
%%-----------------------------
%%
%% The reference list should contain all the references cited in the text, ordered alphabetically by surname (with
%% initials following). If there are several references to the same first author, they should be entered according
%% to the following scheme:
%% 1. One author: chronologically
%% 2. Author, one co-author: alphabetically by co-author, then chronologically
%% 3. Author, two or more co-authors: chronologically.
%%
%% Please note that for papers that have more than five authors, only the first three should be given, followed
%% by "et al."
%%
%% The format for references is the one adopted by A&A (see the example below).
%%
%% To set the reference list in the proper A&A format, we encourage you to use BibTEX and the natbib
%% package instead of the standard 'thebibliography' environment.
%%

%% The following lines are required when using BibTEX (strongly encouraged!):
\bibliographystyle{aa}  % A&A bibliography style file (aa.bst)
\bibliography{sf2a-template} % your references in file: Yourfile.bib

\end{document}